
\normalbaselineskip=20pt
\baselineskip=20pt
\magnification=\magstep1
\hsize 15.0truecm \hoffset 1.0 truecm
\vsize 22.0truecm

\nopagenumbers
\headline={\ifnum \pageno=1
\hfil \else\hss\tenrm\folio\hss\fi}
\pageno=1

\def\lsim{\mathrel{\rlap{\lower4pt\hbox{\hskip1pt$\sim$}}
    \raise1pt\hbox{$<$}}}     
\def\gsim{\mathrel{\rlap{\lower4pt\hbox{\hskip1pt$\sim$}}
    \raise1pt\hbox{$>$}}}   

\hfill IUHET 293

\hfill IU/NTC 95-03

\hfill DFTT--95/7

\hfill January 31, 1995

\centerline{\bf Predicting  the masses of baryons
containing one or two heavy quarks}
\vskip 36pt
\centerline{R. Roncaglia and  D. B. Lichtenberg}

\centerline{\it Department of Physics,
Indiana University, Bloomington, Indiana, 47405}
\centerline{E. Predazzi}

\centerline{Dipartimento di Fisica Teorica,
Universit\`a di Torino and}
\centerline{Istituto Nazionale di Fisica Nucleare,
Sezione di Torino, I-10125, Torino, Italy}

\vskip 2cm

\item{}
The Feynman-Hellmann theorem and semiempirical
mass formulas are used to predict the masses of
baryons containing one or two heavy quarks. In
particular, the mass of the $\Lambda_b$ is predicted
to be $5620 \pm 40$ MeV, a value consistent with
measurements.

\vskip 1cm
\bigskip
\item{} PACS numbers: 12.15.Ff, 12.40.Yx, 14.20.-c, 14.40.-n

\vskip 2cm

\medskip

In a recent paper [1], the Feynman-Hellmann theorem [2,3]
and semiempirical mass formulas [4,5,1] were used as tools
enabling the prediction of the masses
of a number of as yet undiscovered mesons
and baryons. The major point of the paper [1] is
that one can exploit
regularities in the pattern of known hadron masses
to obtain estimates of the masses of unknown
hadrons. Although results are
obtained within the framework of the constituent quark
model, no explicit Hamiltonian is used.
Therefore, the results given in Ref. [1] are complementary
to calculations which use specific potentials
between quarks to calculate hadron masses.

In this brief report, we extend the considerations of
Ref. [1] in order to (A) predict the mass of the
$\Lambda_b$, (B)
make slightly revised predictions of masses of other
baryons containing one heavy quark, and (C) predict
masses of baryons containing two heavy quarks.
A number of authors have recently
considered  baryons [6--12] containing two heavy quarks
in anticipation of future experiments which may
discover these particles. Again, our work is complementary
to that which uses a specific model, and provides
predictions which do not depend upon any explicit
Hamiltonian.

In Ref. [1], it was shown that, in the
constituent quark model,  the Feynman-Hellmann
theorem can be applied to give useful information
about the masses of mesons and baryons even in systems
with relativistic kinematics. Under plausible
assumptions, application of the  Feynman-Hellmann theorem
leads to the conclusion that the energy eigenvalues
(excluding quark masses) of certain mesons and baryons
are smooth, monotonically
decreasing functions of the parameter $\mu$, given by
$$\mu^{-1} = \sum  m_i^{-1},\eqno(1)$$
where the $m_i$ are the masses of the constituent
quarks. These results were found to hold for ground-state
spin-1 mesons and spin-${3\over 2}$ baryons even in
the presence of the colormagnetic (spin-spin)
interaction between pairs of quarks. On the other
hand, the eigenenergies of spin-0 mesons and
spin-${1\over 2}$ baryons do not necessarily
decrease monotonically with increasing $\mu$.

The relation between the eigenenergy $E(12...)$ of
a hadron containing quarks $q_1,q_2,...$
and its mass $M(12...3)$ is
$$M(12...) = E(12...) + \sum_i m_i. \eqno(2)$$
The Feynman-Hellmann theorem is useful in allowing
us to obtain
estimates of $E(12...)$, not $M(12...)$.
Therefore, we can obtain predictions about
hadron masses $M(12...)$  from Eq.\
(2) only if we assume values for the quark masses.

However, we are not allowed to assume arbitrary
values for the quark masses, as we can
obtain constraints on the quark mass
differences from the Feynman-Hellmann
theorem. It was shown in Ref. [1] that the quark
mass differences must satisfy the inequalities
$$m_s-m_q>M(K^*)-M(\rho)=126\pm 4 \ {\rm MeV},\eqno(3)$$
$$m_c-m_s>M(D^*)-M(K^*)=1115\pm 4 \ {\rm MeV},\eqno(4)$$
$$m_b-m_c>M(B^*)-M(D^*)=3316\pm 6 \ {\rm MeV},\eqno(5)$$
where the numbers in (3--5) come from the experimental
values of the vector meson masses as given by
the Particle Data Group [13].

The derivation of (3--5) was given in Ref. [1], but
we briefly repeat it here for (3). From Eq.\ (2), we have
$$M(K^*)= m_q + m_s + E(qs) \eqno(6)$$
and $$M(\rho)= 2m_q + E(qq). \eqno(7)$$
Subtracting (7) from (6) and noting from the
Feynman-Hellmann theorem that $E(qs)< E(qq)$, we
get the inequality in (3).

We can obtain stronger inequalities from spin-${3\over 2}$
baryons in an analogous manner to the derivation
of the inequality (3).
In Ref. [1], the following inequality was derived:
$$m_s-m_q>M(\Sigma^*)-M(\Delta)=153\pm 4 \ {\rm MeV},
\eqno(8)$$
where the numbers are from baryon masses given in
Ref. [13].
This inequality for $m_s-m_q$ is stronger than the
inequality given in (3). However,
the analogous inequalities
$$m_c-m_s>M(\Sigma_c^*)-M(\Sigma^*) \eqno(9)$$
and
$$m_b-m_c>M(\Sigma_b^*)-M(\Sigma_c^*) \eqno(10)$$
do not immediately help us because
the masses of the baryons $\Sigma_c^*$ and $\Sigma_b^*$
are not known experimentally.

However, we can obtain inequalities
for $m_c-m_s$ and $m_b-m_c$ in MeV
from (9) and (10) if we obtain  theoretical estimates of
the masses of $\Sigma_c^*$ and $\Sigma_b^*$. We do this
using a semiempirical mass formula for the spin
splittings among baryons containing a given quark
content [1]. Using the known masses [13] of the
$\Lambda_c$, $\Sigma_c$, and $\Lambda_b$ as input,
we get from the semiempirical formula
 $M(\Sigma_c^*)= 2525$  MeV and $M(\Sigma_b^*)= 5855$ MeV.
Then we obtain the inequalities
$$m_c-m_s>M(\Sigma_c^*)-M(\Sigma^*) = 1140 \pm 20
\ {\rm MeV}, \eqno(11)$$
$$m_b-m_c>M(\Sigma_b^*)-M(\Sigma_c^*) = 3340 \pm 60
\ {\rm MeV}, \eqno(12)$$
where the errors are larger than for the previous
inequalities because of uncertainties
in the semieempirical mass formula [1] and in the
experimental value of the mass of the $\Lambda_b$ [13].

In Ref. [1], the following quark masses
were used for the discussion of baryons:
$m_q=300$ MeV, $ m_s= 475$ MeV, $ m_c= 1640$ MeV,
and $ m_b= 4990$ MeV.
We see that these masses satisfy the inequalities
(8), (11), and (12), and therefore are
potentially good candidates
for use with the Feynman-Hellmann theorem to enable
us to obtain estimates of the masses of baryons with
two heavy quarks. For a detailed discussion of how these
quark masses were arrived at, see Ref.\ [1]. We
use the same masses in this paper except that
we take the mass of the $b$ quark to be
$4985$  MeV,
five MeV lower than the value used in Ref.\ [1].
Thus, we use the quark mass values (in MeV):
$$m_q=300, \quad m_s= 475, \quad m_c= 1640,
\quad m_b= 4985. \eqno(13)$$

The small (5 MeV) difference in the value of $m_b$
comes about as follows: In Ref.\ [1], $m_b$ was obtained
with the mass of the $\Lambda_b$
as input. However, the value of the $\Lambda_b$ mass,
as given in the baryon table
of the Particle Data Group [13], has the
rather large error of $\pm 50$ MeV.
Therefore, in this paper we choose not to use
the $\Lambda_b$ mass as input but to obtain the value
of $m_b$ from the vector mesons. This
revised procedure allows us to  predict the value
of the $\Lambda_b$ mass. We use as input
the quark masses $m_q=300$ MeV, $m_s= 475$ MeV, and
$m_c= 1640$ MeV as found in Ref.\ [1]
for the baryons. We also use as input the observed
masses [13] of the ground-state vector mesons $\rho$,
$K^*$, $\phi$, $D^*$, $D_s^*$, $B^*$, $B_s^*$, $J/\psi$,
and $\Upsilon$. In addition,
we use as input the predicted mass of the $B_c^*$, which
in Ref.\ [1] was found to be 6320 MeV. Then, applying
the Feynman-Hellman theorem as in Ref.\ [1], we
fit the masses of the vector
mesons with a monotonically decreasing function of
the reduced mass $\mu$.
The specific functional form has
no theoretical significance. In practice, we use
a three-parameter exponential curve
similar to that used in Ref.\ [1], but with different
parameters, as the parameters depend on the input
quark masses.  We vary the three parameters of the
curve and the $b$ quark mass so as to get a best
fit to the data. This procedure
yields  $m_b= 4985$ MeV (rounded to the nearest 5 MeV). We
show the fit to the meson eigenenergies in Fig.\ 1.

We obtain predictions for unobserved spin-${3\over 2}$
baryon masses as follows:
We first use as input the observed masses
$\Delta$, $\Sigma^*$, $\Xi^*$, and $\Omega$.
Then we use the semiempirical mass formula of Ref.\ [1]
(with slightly changed parameters because the parameters
depend on the input quark masses) to obtain the masses
of the $\Sigma_c^*$ and $\Xi_c^*$,
using the observed masses of the
$\Lambda_c$, $\Sigma_c$, and $\Xi_c$ as input.
We next subtract the quark  masses of Eq.\ (13)
from the masses of the
spin-${3\over 2}$ baryons $\Delta$, $\Sigma$,
$\Xi^*$, $\Omega$, $\Sigma_c^*$, $\Xi_c^*$
[see Eq.\ (2)] to obtain the corresponding
energy eigenvalues. We then fit a monotonically
decreasing three-parameter
exponential curve through these energies.
The parameters of this curve are slightly different
from the parameters used to fit the baryon
eigenenergies in Ref.\ [1] because we take the $b$
quark mass to be 5 MeV less than in Ref.\ [1].
We then extrapolate the curve to give us the
energy eigenvalues of unknown spin-${3\over 2}$ baryons
including the $\Sigma_b^*$, $\Omega_c^*$, and baryons
containing two heavy quarks. We show this extrapolated
curve in Fig. 2. Because we have to extrapolate rather
a long way in $\mu$, we assign errors of as much as
100 MeV to the eigenergies.

The next step is to
add back the quark masses to obtain the masses of
these spin-${3\over 2}$ baryons. Lastly,
we use the semiempirical
mass formula to obtain the masses of spin-${1\over 2}$
baryons, including the $\Lambda_b$, $\Sigma_b$,
$\Omega_c$, and baryons with two heavy quarks.
In Table I we give  the masses
of baryons containing one or two heavy quarks. Masses
of known baryons used as input are marked with $^a$,
and masses already predicted in Ref.\ [1] are marked
with $^b$. In some of the cases marked with $^b$,
our present predictions differ by 10 MeV from those
given in Ref.\ [1]. These differences are well within
our estimated errors.

Note from Table I that we predict that the mass of the
$\Lambda_b$ is $5620\pm 40$ MeV. (In Ref.\ [1]
the observed mass of the $\Lambda_b$ was
used as input.)
We also predict in Table I
that the mass of the $\Omega_c$ is
$2710\pm 30$ MeV. (This is the same mass
found in Ref.\ [1], but here the error is smaller.)
Both these masses are consistent with
the values $5641\pm 50$ MeV and $2710 \pm 5$
MeV respectively given
by the Particle Data Group [13].

The predicted baryon masses given in Table I satisfy
an inequality derived by Bagan et al. [14], which
says that if quarks 1,2, and 3 are ordered according
to increasing mass, then
$$M(123)\leq M(113) + M(112) - M(111). \eqno(14)$$
If we compare the masses of baryons with
two heavy quarks given in  Table I with masses
calculated with a potential model [12], we find that
our predicted masses are from 20 to 130 Mev larger.

Before ending this paper, we say a few words about
meson masses.
In Ref.\ [1], in order to obtain best fits to the
data on mesons and baryons separately, different
input quark masses were used for mesons and baryons.
The quark masses used for mesons were (in MeV):
$$m_q=300, \quad m_s= 440, \quad m_c= 1590,
\quad m_b= 4920. \eqno(15)$$
Comparing these masses with the ones in Eq.\ (13), we
see that the differences range from 0 (for $m_q$)
to 65 MeV (for $m_b$).
Although there is no good theoretical reason why
effective constituent quark masses need to be the
same in mesons and baryons, it is economical
in parameters to be able to use the same masses.

We immediately see that we cannot use the quark masses
of Eq.\ (15)
for the baryons, as these  mass differences do
not satisfy the baryon inequality (8). In confirmation,
we have
verified numerically that we obtain poor agreement
with known baryon masses when using the quark mass
set of Eq.\ (15). We also find that using
an average of the quark masses of Eqs.\ (13) and (15)
is unsatisfactory, although these average masses
do satisfy all the baryon inequalities (8), (11), and
(12). However, as we see from Fig.\ 1, if we use
the quark masses of Eq.\ (13) as input, we get
reasonable agreement with the experimental vector meson
masses [although the fit is not statistically as
good as with the masses of Eq.\ (15)].
As we have stated, we used
the predicted mass of the $B_c^*$ from Ref.\ [1] as input.
However, all other predictions of the masses of
as yet undiscovered mesons
are the same as those given in Ref.\ [1]
within the errors given in that paper. We conclude
that the quark masses of Eq.\ (13) are suitable to
use for both mesons and baryons in applications
involving the Feynman-Hellmann theorem.

This work was begun while one of us (DBL) was visiting
the Department of Theoretical Physics of the University
of Torino, and he is grateful to members of
that department for their kind hospitality. The work was
supported in part by the U. S. Department of Energy,
the U. S. National Science Foundation, and the Italian
National Institute for Nuclear Physics (INFN).


References
\bigskip

\item{[1]} R. Roncaglia, A. Dzierba, D.B. Lichtenberg,
and E. Predazzi, Phys. Rev. D {\bf 51}, 1248 (1995).

\item{[2]} R.~P. Feynman, Phys. Rev. {\bf 56}, 340 (1939).

\item{[3]} H. Hellmann, Acta Physicochimica URSS
I, 6, 913 (1935); IV, 2, 225 (1936); Einf\"uhr\-ung
in die Quantenchemie (F. Deuticke,
Leipzig and Vienna, 1937) p. 286.

\item{[4]} X. Song, Phys.\ Rev.\ D
{\bf 40}, 3655 (1989).

\item{[5]} Yong Wang and D. B. Lichtenberg,
Phys.\  Rev.\ D {\bf 42}, 2404 (1990).

\item{[6]}
M. J. Savage and  M. B. Wise,
Phys. Lett. B {\bf 248}, 177 (1990).

\item{[7]}
M. J. White and M. J. Savage,
Phys. Lett. B {\bf 271}, 410 (1991).

\item{[8]}
E. Bagan, M. Chabab, S. Narison,
Phys. Lett. B {\bf 306}, 350 (1993).

\item{[9]}
A. F. Falk, M. Luke, M. J. Savage,
Phys. Rev. D {\bf 49}, 555 (1994).

\item{[10]}
M. Angel, Phys. Lett. B {\bf 321}, 407 (1994).

\item{[11]}
M. Bander, A. Subbaraman, report UCI-TR-94-30, 1994
(unpublished).

\item{[12]}
J. M. Richard, Joseph Fourier University
(Grenoble) preprint, 1994 (unpublished).

\item{[13]} Particle Data Group: L. Montanet et al.,
Phys. Rev. D {\bf 50}, 1171 (1994).

\item{[14]} E. Bagan, H. G. Dosch, P. Gosdzinsky,
S. Narison, and J. M. Richard, Z. Phys. C {\bf 64}, 57
(1994). There is a misprint in Eq.\ (33) of this
paper. The symbol $\ge$ should be replaced by $\le$.

\vfill\eject

TABLE I. Masses of
baryons containing one or  two heavy quarks.
In column 3 we show  predictions
for ground-state spin-1/2 baryons (denoted by $M_A$ in the
first row of this column) whose first two quarks
have an antisymmetric spin wave function.
In column 4 we show predictions for ground-state
spin-1/2 baryons (denoted by $M_S$)
with symmetric spin wave functions in the first two quarks.
In column 5 we show predictions for the ground-state
spin-3/2 baryons (denoted by $M^*$).

$$\vbox {\halign {\hfil #\hfil &&\quad \hfil #\hfil \cr
\cr \noalign{\hrule}
\cr \noalign{\hrule}
\cr
Name & Quark content & $M_A$ (MeV) & $M_S$ (MeV) &
$M^*$ (MeV) \cr
\cr \noalign{\hrule}
\cr
$\Lambda_c, \Sigma_c, \Sigma_c^*$ & $qqc$ &
$2285\pm 1 ^a$ & $2453\pm 3^a$ & $2520\pm 20$ \cr
$\Xi_c,\Xi_c',\Xi_c^*$ &$qsc$ & $2468\pm 3^a$ &
$2580\pm 20$ & $2650\pm 20$ \cr
$\Omega_c, \Omega_c^*$ &$ssc$ & ---    & $2710\pm 30^{b,c}$
& $2770\pm 30^b$ \cr
$\Lambda_b, \Sigma_b, \Sigma_b^*$ & $qqb$ &
$5620\pm 40$ & $5820\pm 40^b$ & $5850\pm 40^b$
\cr
$\Xi_b,\Xi_b',\Xi_b^*$ &$qsb$ & $~5810\pm 40^b$ &
$5950\pm 40^b$ & $5980\pm 40^b$ \cr
$\Omega_b,\Omega_b^*$ & $ssb$ & ---   & $6060\pm 50^b$ &
$6090\pm 50^b$ \cr
$\Xi_{cc},\Xi_{cc}^*$ &$ccq$ & --- &
$3660\pm 70$ & $3740\pm 70$ \cr
$\Omega_{cc}, \Omega_{cc}^*$ & $ccs$
& ---   & $3740\pm 80$ & $3820\pm 80$ \cr
$\Xi_{cb},\Xi_{cb}',\Xi_{cb}^*$ &$qcb$ & $6990\pm 90$ &
$7040\pm 90$ & $7060\pm 90$ \cr
$\Omega_{cb}, \Omega_{cb}',\Omega_{cb}^*$ & $scb$
& $7060\pm 90$  & $7090\pm 90$ & $7120\pm 90$ \cr
$\Xi_{bb},,\Xi_{bb}^*$ &$bbq$ & --- &
$10340\pm 100$ & $10370\pm 100$ \cr
$\Omega_{bb}, \Omega_{bb}^*$ & $bbs$
& ---  & $10370\pm 100$ &$10400\pm 100$ \cr

\cr \noalign{\hrule}
\cr \noalign{\hrule}
}}$$

$^a$Masses of known baryons used as input.

$^b$Masses also predicted in Ref.\ [1]. The masses in this
table sometimes differ from the masses of Ref.\ [1]
by up to 10 MeV because of slightly different input
parameters. These differences are well within our
estimated errors.

$^c$This prediction agrees with the value
$2710\pm 5$ MeV given in the
full listings of the Particle Data Group [13]. We did
not use the Particle Data Group
value as input because the $\Omega_c$
was omitted from their summary baryon table.
\vskip 1cm

Figure captions
\bigskip

FIG.\ 1. Fit to the energy eigenvalues of vector
mesons with an exponential curve.
Open circles denote mesons whose
eigenergies are obtained from experiment by subtracting
the quark masses of Eq.\ (13), and the solid circle denotes
an eigenenergy obtained from the mass of the
$B_c^*$ predicted in Ref.\ [1].

\bigskip

FIG.\ 2. Energy eigenvalues of baryons with spin
${3\over 2}$. Open circles denote
baryons whose eigenergies are obtained
from experiment [13] by subtracting the quark masses of
Eq. (13), triangles denote eigenergies obtained with
the aid of the baryon semiempirical mass formula of Ref.
[1], and solid circles denote
predicted eigenergies obtained from
an exponential curve by extrapolation.

\bye